\documentclass[12pt,preprint]{aastex}
\usepackage{amssymb,multirow,color}
\makeatletter
\renewcommand{\vec}[1]{\mathbf{#1}}

\renewcommand{\(}{\left(}
\renewcommand{\)}{\right)}

\begin{document}
\title{Extended Scaling Laws in Numerical Simulations of MHD Turbulence}
\author{Joanne Mason$^1$, Jean Carlos Perez$^{2,3}$, Fausto Cattaneo$^1$, Stanislav Boldyrev$^2$} 
\affil{${~}^1$Department of Astronomy and Astrophysics, University of Chicago, 
5640 S. Ellis Ave, Chicago, IL 60637\\
${~}^2$Department of Physics, University of Wisconsin at Madison, 1150 University Ave, 
Madison, WI 53706 \\
${~}^3$Space Science Center and Department of Physics, University of New Hampshire, Durham, NH 03824 \\ {\sf jmason@flash.uchicago.edu}, {\sf jcperez@wisc.edu}, {\sf cattaneo@flash.uchicago.edu}, {\sf boldyrev@wisc.edu} }
\date{\today}

\begin{abstract}
Magnetised turbulence is ubiquitous in astrophysical systems, where it notoriously spans a broad range of spatial scales. Phenomenological theories of MHD turbulence describe the self-similar dynamics of turbulent fluctuations in the inertial range of scales. Numerical simulations serve to guide and test these theories. However, the computational power that is currently available restricts the simulations to Reynolds numbers that are significantly smaller than those in astrophysical settings. In order to increase computational efficiency and, therefore, probe a larger range of scales, one often takes into account the fundamental anisotropy of field-guided MHD turbulence, with gradients being much slower in the field-parallel direction. The simulations are then optimised by employing the reduced MHD equations and relaxing the field-parallel numerical resolution. In this work we explore a different possibility. We propose that there exist certain quantities that are remarkably stable with respect to the Reynolds number. As an illustration, we study the alignment angle between the magnetic and velocity fluctuations in MHD turbulence, measured as the ratio of two specially constructed structure functions. We find that the scaling of this ratio can be extended surprisingly well into the regime of relatively low Reynolds number. However, the extended scaling becomes easily spoiled when the dissipation range in the simulations is under-resolved. Thus, taking the numerical optimisation methods too far can lead to spurious numerical effects and erroneous representation of the physics of MHD turbulence, which in turn can affect our ability to correctly identify the physical mechanisms that are operating astrophysical systems.
\end{abstract}


\maketitle

\section{Introduction}

Magnetised turbulence pervades the universe. It is likely to play an important role in the transport of energy, momentum and charged particles in a diverse range of astrophysical plasmas. It is studied with regards to its influence on the generation of magnetic fields in stellar and planetary interiors, small-scale structure and heating of stellar winds, the transport of angular momentum in accretion discs,  gravitational collapse and star formation in molecular clouds, the propagation and acceleration of cosmic rays, and interstellar scintillation \citep[e.g.,][]{biskamp_03,kulsrud_04,mckee_o07,goldstein_rm95,brandenburg_n11,schekochihin_c07}. The effects of magnetised turbulence need to be taken into account when analysing astrophysical observations and also when modelling astrophysical processes. 

The simplest theoretical framework that describes magnetised plasma turbulence is that of incompressible magnetohydrodynamics (MHD),  
  \begin{equation}
 \label{eq:mhd-elsasser}
  \(\frac{\partial}{\partial t}\mp\vec V_A\cdot\nabla\)\vec
  z^\pm+\left(\vec z^\mp\cdot\nabla\right)\vec z^\pm = -\nabla P + \nu \nabla^2 \vec z^{\pm}+\vec f^\pm, 
   \end{equation}
   \begin{equation}
   \label{eq:div}
   \nabla \cdot {\bf z}^{\pm}=0, 
  \end{equation}
where the Els\"asser variables are defined as $\vec z^\pm=\vec v\pm\vec
b$, $\vec v$ is the fluctuating plasma velocity, $\vec b$ is the
fluctuating magnetic field normalized by $\sqrt{4 \pi \rho_0}$, ${\bf
V}_A={\bf B}_0/\sqrt{4\pi \rho_0}$ is the Alfv\'en velocity based upon the uniform
background magnetic field ${\vec B_0}$, $P=(p/\rho_0+b^2/2)$, $p$ is the plasma pressure,
$\rho_0$ is the background plasma density, $\vec f^\pm$ represents forces that drive the 
turbulence at large scales and for simplicity we have taken the case in which the fluid viscosity $\nu$ is equal to the magnetic resistivity. Energy is transferred to smaller scales by the nonlinear interactions of oppositely propagating Alfv\'en wavepackets \citep{kraichnan_65}. This can be inferred directly from equation (\ref{eq:mhd-elsasser}) by noting that in the absence of forcing and dissipation,  if ${\bf z}^\mp({\bf x},t)\equiv 0$ then any function ${\bf z}^\pm({\bf x},t)=F^\pm({\bf
x}\pm {\bf V}_A t)$ is an exact nonlinear solution that propagates parallel and anti-parallel to $\vec{B_0}$ with the Alfv\'en speed.  The efficiency of the nonlinear interactions splits MHD turbulence into two regimes. The regime in which the linear terms dominate over the nonlinear terms is known as `weak' MHD turbulence, otherwise the turbulence is `strong'. In fact, it has been demonstrated both analytically and numerically that the MHD energy cascade occurs predominantly in the plane perpendicular to the guiding magnetic field. This ensures that even if the turbulence is weak at large scales it encounters the strong regime as the cascade proceeds to smaller scales. MHD turbulence in astrophysical systems is therefore typically strong. 

For strong MHD turbulence, \citet{goldreich_s95} argued that the linear and nonlinear terms in equations (\ref{eq:mhd-elsasser}) should be approximately balanced at all scales, known as the critical balance condition. Consequently, \cite{goldreich_s95} postulated that the wave packets get progressively
elongated in the direction of the guide field as their scale decreases (with the field-parallel lengthscale $l$ and field-perpendicular scale $\lambda$ related by~$l\propto \lambda^{2/3}$) and that the field-perpendicular energy spectrum takes the Kolmogorov form
$E_{GS}(k_\perp)\propto k_\perp^{-5/3}$. 

Recent high resolution direct numerical simulations with a strong guide field ($v_A\geq
5v_{rms}$) do indeed verify the strong anisotropy of the turbulent fluctuations,
however, the field-perpendicular energy
spectrum appears to be closer to $E(k_\perp)\propto k_\perp^{-3/2}$
\citep[e.g.,][]{maron_g01,muller_g05,mason_cb08,perez_b09,perez_b10}.  
A resolution to this contradiction was proposed in \cite{boldyrev_06}. Therein it was
suggested that in addition to the elongation of the eddies in the
direction of the guiding field, the fluctuating velocity and magnetic
fields at a scale $\lambda\sim 1/k_{\perp}$ are aligned within a small
scale-dependent angle in the field perpendicular plane, $\theta_\lambda
\propto \lambda ^{1/4}$. In this model the wavepackets are three-dimensionally anisotropic. Scale-dependent dynamic
alignment reduces the strength of the nonlinear interactions and leads
to the field-perpendicular energy spectrum $E(k_{\perp}) \propto k_{\perp}^{-3/2}$.

Although the two spectral exponents $-5/3$ and $-3/2$ are close together in numerical value, the physics of the energy cascade in each model is different. The difference between the two exponents is especially important for inferring the behaviour of processes in astrophysical systems with extended inertial intervals. For example, the two exponents can lead to noticeably different predictions for the rate of turbulent heating in coronal holes and the solar wind \citep[e.g.,][]{chandran_etal10,chandran10,podesta_b10}. Thus, there is much interest in accurately determining the spectral slope from numerical simulations. Unfortunately, the Reynolds numbers that are currently accessible by most direct numerical simulations do not exceed a few thousand, which complicates the precise identification of scaling exponents. Techniques for careful optimisation of the numerical setup and alternative ways of differentiating between the competing theories are therefore much sought after.

Maximising the extent of the inertial range is often achieved by implementing physically motivated simplifying assumptions.  For example, since the turbulent cascade proceeds predominantly in the field-perpendicular plane it is thought that the shear-Alfv\'en waves control the dynamics while the pseudo-Alfv\'en waves play a passive role (see, e.g., \cite{maron_g01}). If one neglects the pseudo-Alfv\'en waves (i.e. removes the fluctuations parallel to the strong guide field) one obtains a system that is equivalent to the reduced MHD system (RMHD) that was originally derived in the context of fusion devices by \cite{kadomtsev_p74} and \cite{strauss_76} (see also \cite{biskamp_03,oughton_dm04}). Incompressibility then enables the system to be further reduced to a set of two scalar equations for the Els\"asser potentials, resulting in a saving of approximately a factor of two in computational costs. Further computational savings can be made by making use of the fact that the wavepackets are elongated. Hence variations in the field-parallel direction are slower than in the field-perpendicular plane and a reduction in the field-parallel resolution would seem possible. Indeed, this is widely used as an optimisation tool in numerical simulations of the inertial range of field-guided MHD turbulence \cite[e.g.,][]{maron_g01,mason_cb08, grappin_m10, perez_b10}. The accumulated computational savings can then be re-invested in reaching larger Reynolds numbers for the field-perpendicular dynamics. 

Additionally, it is advantageous to seek other ways of probing the universal scaling of MHD turbulence. In this work we point out a rather powerful method, which is based on the fact that there may exist certain quantities in MHD turbulence that exhibit very good scaling laws even for turbulence with relatively low Reynolds numbers. The situation here is reminiscent of  the well known phenomenon of extended self-similarity in hydrodynamic turbulence \citep{benzi_etal93}. We propose that one such ``stable'' object is the alignment angle between the velocity and magnetic fluctuations, which we measure as the ratio of two specially constructed structure functions. This ratio has been recently measured in numerical simulations in an attempt to differentiate among various theoretical predictions (\cite{beresnyak_l06, mason_cb06}). Also, it has recently been shown by  \cite{podesta_etal09} that the same measurement is accessible through direct observations of solar wind turbulence. Scale-dependent alignment therefore has practical value: its measurement may provide an additional way of extracting information about the physics of the turbulent cascade from astrophysical observations. In the present work we conduct a series of numerical simulations with varying resolutions and Reynolds numbers. We find that as long as the simulations are well resolved, the alignment angle exhibits a universal scaling behavior that is virtually independent of the Reynolds number of the turbulence. Moreover, we find that the {\em length} of scaling range for this quantity extends to the smallest resolved scale, independently of the Reynolds number. This means that although the dissipation spoils the power-law scaling behaviour of each of the structure functions, the dissipation effects cancel when the ratio of the two functions is computed and the universal inertial-range scaling extends deep in the dissipation region.  

The described method allows the inference of valuable scaling laws from numerical simulations, experiments, or observations of MHD turbulence with limited Reynolds number. However, one can ask how well the extended-scaling method can be combined with the previously mentioned optimisation methods relying on the reduced MHD equations and a decreased parallel resolution. We check that reduced MHD does not alter the result. However, when the dissipation region becomes under-resolved (as can happen, for example, when the field-parallel resolution is decreased), the extended scaling of the alignment angle deteriorates significantly. Thus the optimisation technique that works well for viewing the inertial range of the energy spectra should not be used in conjunction with the extended-scaling measurements that probe deep into the dissipation region. The remainder of this paper will report the findings of a series of numerical measurements of the alignment angle in simulations with different Reynolds numbers and different field-parallel resolutions in both the MHD and RMHD regimes. The aim is to address the need to find an optimal numerical setting for studying strong MHD turbulence and to raise caution with regards to the effects that implementing simplifying assumptions in the numerics can have on the solution and its physical interpretation.

\section{Numerical results}
We simulate driven incompressible magnetohydrodynamic turbulence in the presence of a strong uniform background magnetic field, $\vec{B_0}=B_0\vec{\hat e_z}$. The MHD code solves equations (\ref{eq:mhd-elsasser},\ref{eq:div}) on a periodic, rectangular domain with aspect ratio $L_{\perp}^2 \times L_\|$, where the subscripts denote the directions perpendicular and parallel to $\vec{B_0}$ and we take $L_{\perp}=2\pi$. A fully dealiased 3D pseudospectral algorithm is used to perform the spatial discretisation on a grid with a resolution of
 $N_{\perp}^2\times N_\|$ mesh points. The RMHD code solves the reduced MHD counterpart to equations (\ref{eq:mhd-elsasser},\ref{eq:div}) in which $\vec z^\pm=(z_x^\pm, z_y^{\pm},0)$ (see \cite{perez_b08}).
 
The domain is elongated in the direction of the guide field in order to accommodate the elongated wavepackets and to enable us to drive the turbulence in the strong regime while maintaining an inertial range that is as extended as possible (see \cite{perez_b10}). 
The random forces are applied in Fourier
space at wavenumbers $2\pi/L_{\perp} \leq k_{\perp} \leq 2 (2\pi/L_{\perp})$, $(2\pi/L_\|) \leq
k_\|\leq (2\pi/L_\|)n_\|$, where we shall take $n_\|=1$ or $n_\|=2$. The forces have no component along $z$ and are solenoidal in the $xy$-plane. 
All of the Fourier coefficients outside the above range of wavenumbers are zero and inside that range are
Gaussian random numbers with amplitudes chosen so that $v_{rms}\sim 1$. The individual random values are refreshed independently on average every $\tau=n_{\tau} L_{\perp}/2\pi v_{rms}$, i.e. the force is updated approximately $1/n_{\tau}$ times per turnover of the large-scale eddies. The variances $\sigma_{\pm}^2=\langle |\vec f^{\pm} |^2\rangle$ control the average rates of energy injection into the $z^+$ and $z^-$ fields.   
The results reported in this paper are for the balanced case $\sigma_+\approx \sigma_-$. 
In all of the simulations
performed in this work we will set the background magnetic field $B_0=5$ in velocity $rms$ units. Time is normalised to the large
scale eddy turnover time $\tau_0=L_\perp/2\pi v_{rms}$. The field-perpendicular Reynolds
number is defined as $Re_{\perp}=v_{rms}(L_\perp/2\pi)/\nu$.

The system is evolved until a stationary state is reached, which is confirmed by observing the time evolution of the total energy of the fluctuations, and the data are then sampled in intervals of the order of the eddy turnover time. All results presented correspond to averages over approximately 30 samples. We conduct a number of MHD and RMHD simulations with different resolutions, Reynolds numbers and field-parallel box sizes. The parameters for each of the simulations are shown in Table~\ref{tab:params}.

\begin{deluxetable}{ccccccc} 
\tablecolumns{7} 
\tablewidth{0pc} 
\tablecaption{Simulation Parameters} 
\tablehead{ \colhead{Case} & \colhead{Regime}   & \colhead{$N_{\perp}$} & \colhead{$N_{\|}$}    & \colhead{$L_{\|}/L_{\perp} $} & \colhead{$Re_{\perp}$} & \colhead{$n_{\tau}$}}
\startdata 
M1 & MHD & 256 & 256 & 5 & 800 & 1  \\
M2 & MHD & 512 & 512 & 5 & 2200 & 1  \\
M3 & MHD & 512 & 512 & 5 & 2200 & 0.1  \\ 
M4 & MHD & 512 &512 & 10 & 2200 & 0.1 \\ 
M5 & MHD & 512 & 256 & 10 & 2200 &0.1 \\  
R1 & RMHD & 512 & 512 & 6 & 960 & 0.1 \\ 
R2 & RMHD  & 512 & 512 & 6 & 1800 & 0.1 \\ 
R3 & RMHD & 256 & 256 & 6 & 960 &0.1 \\ 
\enddata 
\label{tab:params}
\end{deluxetable} 

For each simulation we calculate the scale-dependent alignment angle 
between the shear-Alfv\'en velocity and magnetic field fluctuations. We therefore define velocity and magnetic differences as
$\delta {\bf v}_{r}={\bf v}({\bf x}+{\bf r})-{\bf v}({\bf x})$ and $\delta {\bf b}_{r}={\bf b}({\bf x}+{\bf r})-{\bf b}({\bf x})$, 
where ${\bf r}$ is a  point-separation vector in the plane perpendicular to  ${\bf B}_0$. In the MHD case the pseudo-Alfv\'en fluctuations are removed by subtracting the component that is parallel to the local guide field, i.e. we construct $\delta {\tilde {\bf v}}_{r}=\delta {\bf v}_{r}-(\delta 
{\bf v}_{r}\cdot {\bf n}){\bf n}$ (and similarly for $\delta {\tilde {\bf b}}_{r}$) where ${\bf n}={\bf B}({\bf x})/\vert {\bf B}({\bf x})\vert$. In the RMHD case fluctuations parallel to $\vec{B_0}$ are not permitted and hence the projection is not necessary. We then measure the ratio of the second order structure functions 
\begin{equation}
\label{eq:angle1}
\frac{S^{v \times b}_r}{S^{vb}_r}=\frac{\langle | \mathbf{\delta \tilde v_r} \times \mathbf{\delta \tilde b_r} | \rangle }{\langle | \mathbf{\delta \tilde v_r}| |\mathbf{\delta \tilde b_r}| \rangle}
\end{equation}
where the average is taken over different positions of the point $\vec x$ in a given field-perpendicular plane, over all such planes in the data cube, and then over all data cubes.
By definition of the cross product
\begin{equation}
\label{eq:angle2}
\frac{S^{v \times b}_r}{S^{vb}_r} \approx \sin(\theta_r)\approx \theta_r,
\end{equation}
where $\theta_r$ is the angle between $\delta \tilde \vec v_r$ and $\delta \tilde \vec b_r$ and the last approximation is valid for small angles. We recall that the theoretical prediction is $\theta_r \propto r^{1/4}$ \citep{boldyrev_06}.

\begin{figure}[!tb]
\resizebox{1.\textwidth}{!}{\includegraphics{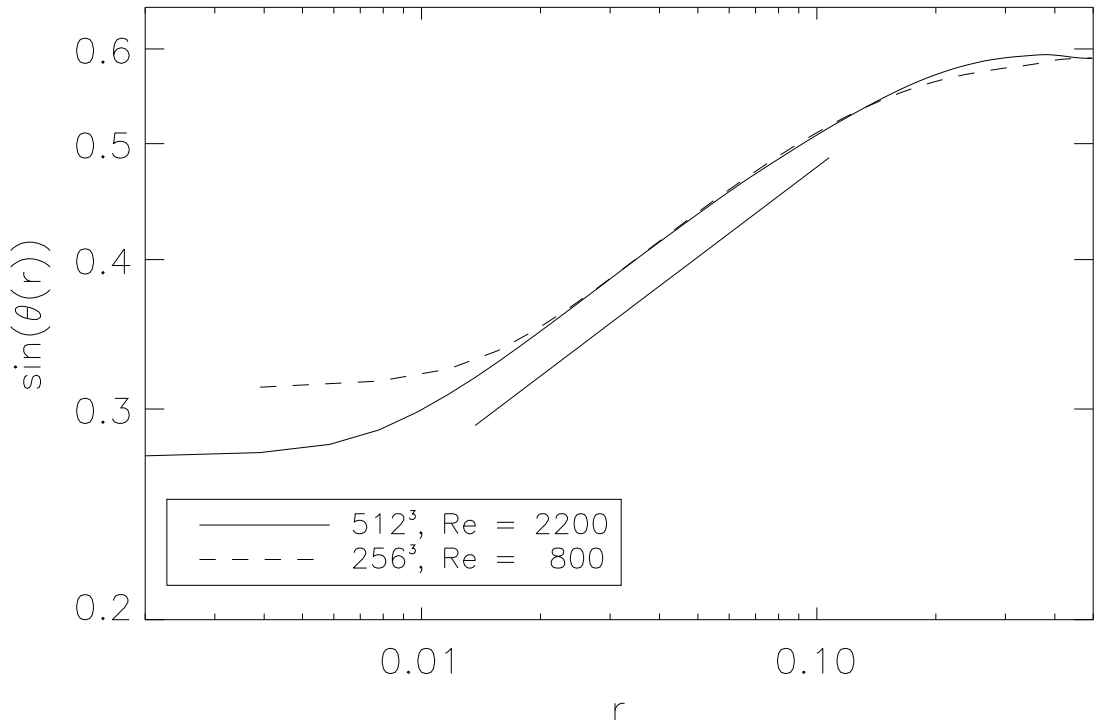}}
    \caption{The alignment angle for two MHD simulations (M1 and M2) corresponding to doubling the resolution and increasing the Reynolds number. In all the figures, the straight line has a slope of $1/4$.}\label{fig:angle_n_re}
\end{figure}

\begin{figure} [tbp]
\resizebox{1.\textwidth}{!}{\includegraphics{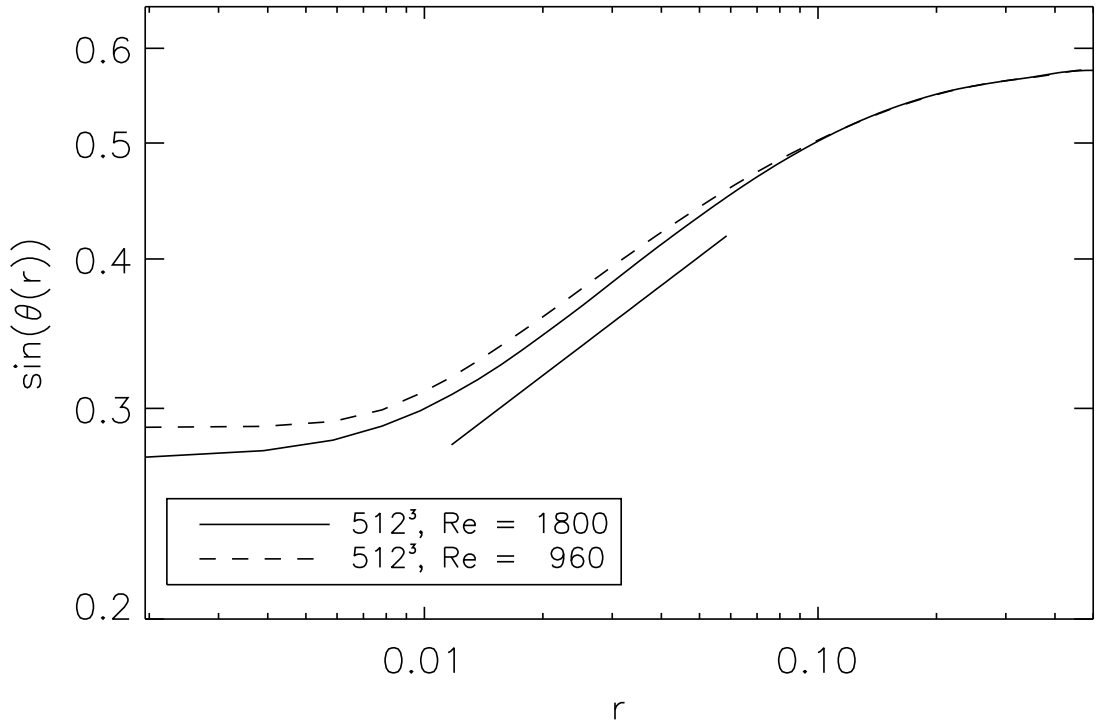}}
\caption{The alignment angle for two RMHD simulations (R1 \& R2) with the same resolution but different Reynolds number.}
\label{fig:angle_re}
\end{figure}

\begin{figure} [tbp]
\resizebox{1.\textwidth}{!}{\includegraphics{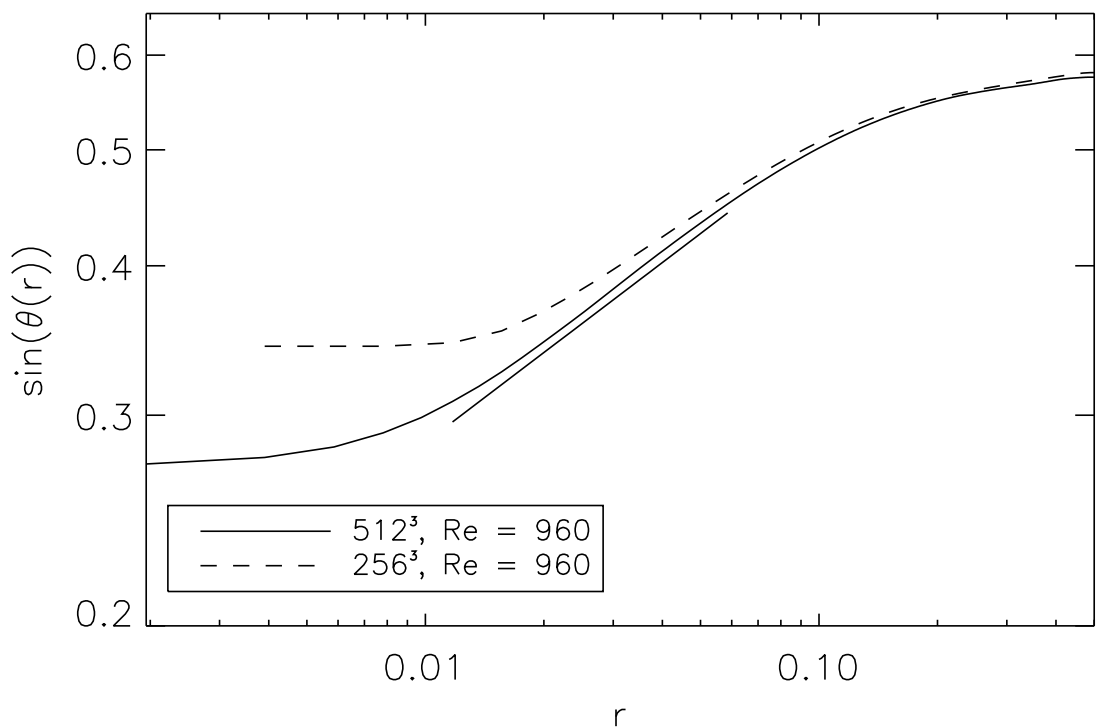}}
\caption{The alignment angle for two RMHD simulations (R2 \& R3) with the same Reynolds number but different resolution.}
\label{fig:angle_n}
\end{figure}

Figure~\ref{fig:angle_n_re} illustrates the ratio (\ref{eq:angle2}) as a function of the separation $r=|\vec{r}|$ for two MHD simulations (M1 and M2) corresponding to a doubling of the resolution from $256^3$ to $512^3$ mesh points with the Reynolds number increased from $Re_{\perp}=800$ to $2200$. Excellent agreement with the theoretical prediction $\theta \propto r^{1/4}$ is seen in both cases. As the resolution and Reynolds number increase, the scale-dependence of the alignment angle persists to smaller scales. Indeed, we believe that the point at which the alignment saturates can be identified as the dealiasing scale, $k_d=N/3=85,170$ corresponding in configuration space to $r_d \approx 1/2k_d \approx 0.006, 0.003$ for the $256^3,512^3$ simulations, respectively. This is verified in Figure~\ref{fig:angle_re} that shows that alignment is largely insensitive to the Reynolds number (provided that the system is turbulent) and Figure~\ref{fig:angle_n} that shows that the saturation point decreases by a factor of approximately 2 as the resolution doubles at fixed Reynolds number. Thus as computational power increases, allowing higher resolution simulations to be conducted, we expect to find that scale-dependent alignment persists to smaller and smaller scales. 

The fact that even in the lower Reynolds number cases scale-dependent alignment is clearly seen over quite a wide range of scales is particularly interesting, as in those cases only a very short inertial range can be identified in the field-perpendicular energy spectrum, making the identification of spectral exponents difficult (see Figure~1 in \cite{mason_cb06}).  In the larger $Re$ cases, we can estimate the inertial range of scales in configuration space to be the range of $r \sim 1/2k$ over which the energy spectrum displays a power law dependence. The field-perpendicular energy spectrum for the Case M2 is shown in Figure~1 in \cite{mason_cb08}, with the inertial range corresponding to approximately $4 \lesssim k \lesssim 20$,  i.e.  $0.025 \lesssim r \lesssim 0.125$. Comparison with Figure~\ref{fig:angle_n_re} shows that a significant fraction of the region over which the scaling $\theta_r \propto r^{1/4}$ is observed corresponds to the dissipative region, i.e. that ratios of structure functions appear to probe deeper than the inertial range that is suggested by the energy spectra. 

\begin{figure}[!tb]
\resizebox{1.\textwidth}{!}{\includegraphics{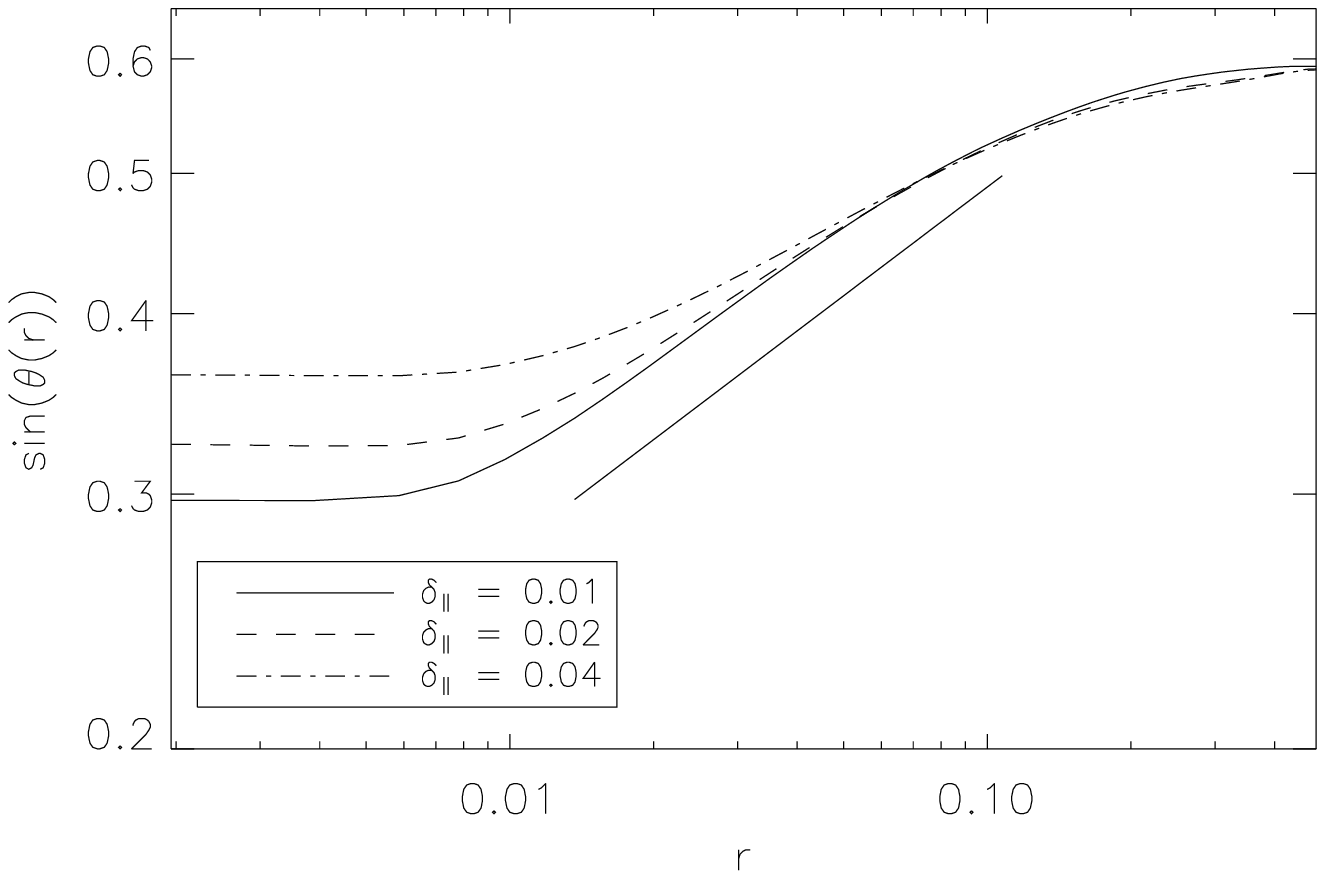}}
    \caption{The alignment angle for three MHD simulations (M3, M4 \& M5) with different field-parallel resolutions $\delta _\|= (L_\|/2\pi)/N_\|$ and $Re=2200$.}\label{fig:angle_nz}
\end{figure}

We now consider the effect on the alignment ratio of decreasing the field-parallel resolution. Figure~\ref{fig:angle_nz} shows the results from three MHD simulations (M3, M4 \& M5) for which the field-parallel resolution decreases by a factor of two, twice. As the resolution decreases the extent of the self-similar region diminishes and the scale-dependence of the alignment angle becomes shallower. If one were to calculate the slope for the lowest field-parallel resolution case (M5) one would find a scale-dependence that is shallower than the predicted power law exponent of $1/4$. This may lead one to conclude (incorrectly) that scale-dependent alignment is not a universal phenomenon in MHD turbulence. However, the effect is obviously a result of the poor resolution rather than being an attribute of the alignment mechanism itself. 

Finally, we mention that for the three cases illustrated in Figure~\ref{fig:angle_nz}, the field-perpendicular energy spectra (not shown) display no appreciable difference. Since the Reynolds number is moderate the inertial range in $k$-space is quite short. However, when the spectra are compensated with $k^{3/2}$ and $k^{5/3}$ the former results in a better fit in all cases. This happens for two reasons. 
First, the stronger deviation from the alignment scaling  $\theta \propto r^{1/4}$ occurs deeper in the dissipation region, that is, further from the inertial interval where the energy spectrum is measured. Second, according to the relationship between the scaling of the alignment angle and the energy spectrum, a noticeable change in the scaling of the alignment angle leads to a relatively small change in the scaling of the field-perpendicular energy spectrum.

\section{Conclusion}

There are two main conclusions that can be drawn from our results. The first is that the measurement of the alignment angle, which is composed of the ratio of two structure functions, appears to display a self-similar region of significant extent, even in the moderate Reynolds number case which requires only a moderate resolution. We have checked that plotting the numerator and the denominator of the alignment ratio separately as functions of the increment $r$ displays only a very limited self-similar region, from which scaling laws cannot be determined. A clear scaling behaviour is also not found when one plots the numerator versus the denominator as is the case in extended self-similarity \citep{benzi_etal93}. The result is interesting in its own right. It also has important practical value as it allows us to differentiate effectively between competing phenomenological theories through numerical simulations conducted in much less extreme parameter regimes than would otherwise be necessary. The result could be especially useful if it extends to ratios of structure functions for which an exact relation, such as the \citet{politano_p98b} relations, is known for one part, as it would then allow the inference of the scaling of the other structure function. Reaching a consensus on the theoretical description of magnetised fluctuations in the idealised incompressible MHD system represents the first step towards the ultimate goal of building a theoretical foundation for astrophysical turbulence.

The second main result that can be drawn from our work is that the measurement of the alignment angle appears to probe deep into the dissipation region and hence it is necessary to adequately resolve the small scale physics. As the field-parallel resolution is decreased, numerical errors contaminate the physics of the dissipative range and affect measurement of the alignment angle. As the decrease in resolution is taken to the extreme, the errors propagate to larger scales and may ultimately spoil an inertial range of limited extent. We propose that similar contamination effects should also arise through any mechanism that has detrimental effects on the dissipative physics. Mechanisms could include pushing the Reynolds number to the extreme or using hyperdiffusive effects. For example, our results may provide an explanation for the numerical findings by \cite{beresnyak11} who noticed a flattening of the alignment angle in simulations of MHD turbulence with a reduced parallel resolution and strong hyperdiffusivity.

We also point out that the result recalls  the phenomenon of extended self-similarity in isotropic hydrodynamic turbulence \citep{benzi_etal93}, which refers to the extended self-similar region that is found when one plots one structure function versus another, rather than as a function of the increment. Our finding is fundamentally different however, in the sense that the self-similar region only becomes apparent when one plots ratios of structure functions versus the increment, rather than structure functions versus other structure functions. 
Our result appears to be due to a non-universal features in the amplitudes  of the functions, rather than their arguments, cancelling when the ratios are plotted.  Whether such a property holds for other structure functions in MHD turbulence is an open and intriguing question. This is a subject for our future work.

\acknowledgments
We would like to thank Leonid Malyshkin for many helpful discussions.
This work was supported by the NSF Center for Magnetic
Self-Organization in Laboratory and Astrophysical Plasmas
at the University of Chicago and the University of Wisconsin - Madison, the US DoE awards DE-FG02-07ER54932, DE-SC0003888, DE-SC0001794, and the NSF grants PHY-0903872 and AGS-1003451. This research used resources of the 
Argonne Leadership Computing Facility at Argonne National Laboratory, which is supported by the Office of 
Science of the U.S. Department of Energy under contract DE-AC02-06CH11357.


\begin{thebibliography}{99}

\bibitem[Benzi et al.(1993)]{benzi_etal93} Benzi, R., Ciliberto, S., Tripiccione, R., Baudet, C., Massaioli, F. \& Succi, S. 1993, \pre, 48, R29
\bibitem[Beresnyak(2011)]{beresnyak11} Beresnyak, A.  2011, \prl, 106, 075001
\bibitem[Beresnyak \& Lazarian(2006)]{beresnyak_l06} Beresnyak, A. \& Lazarian, A. 2006, \apj, 640, L175
\bibitem[Biskamp(2003)]{biskamp_03} Biskamp, D. 2003,  Magnetohydrodynamic Turbulence (Cambridge University Press, Cambridge)
\bibitem[Boldyrev(2006)]{boldyrev_06} Boldyrev, S. 2006, \prl, 96, 115002
  \bibitem[Boldyrev et al.(2009)]{boldyrev_mc09} Boldyrev, S., Mason, J. \& Cattaneo, F. 2009, \apj, 699, L39.
\bibitem[Brandenburg \& Nordlund(2011)]{brandenburg_n11} Brandenburg, A. \& Nordlund A. 2011, Rep. Prog. Phys., 74, 046901
\bibitem[Chandran et al.(2010)]{chandran_etal10}  
	Chandran, B. D. G., Li, B., Rogers, B. N., Quataert, E., \& Germaschewski, K. 2010, \apj, 720, 503
\bibitem[Chandran(2010)]{chandran10} Chandran, B. D. G. 2010, \apj, 720, 548
\bibitem[Goldreich \& Sridhar(1995)]{goldreich_s95} Goldreich,
  P. \& Sridhar, S. 1995, \apj, 438, 763
\bibitem[Goldstein, Roberts \& Matthaeus(1995)]{goldstein_rm95} Goldstein, M.~L., Roberts, D.~A., \& Matthaeus, W.~H. 1995, Ann. Rev. Astron. Astrophys., 33, 283
\bibitem[Grappin \& M\"uller(2010)]{grappin_m10} 
	Grappin, R. \&  M\"uller, W.-C. 2010, \pre, 82, 026406
\bibitem[Kadomtsev \& Pogutse(1974)]{kadomtsev_p74} Kadomtsev,
  B. B. \& Pogutse, O. P. 1974, Sov. Phys. JETP, 38, 283
\bibitem[Kolmogorov(1941)]{kolmogorov_41} Kolmogorov, A.N. 1941, Dokl. Akad. Nauk SSSR, 32, 16 (reprinted in Proc. R. Soc. Lond. A, 434, (1991) 15)
\bibitem[Kraichnan(1965)]{kraichnan_65} Kraichnan, R. H. 1965, Phys. Fluids, 8, 1385
\bibitem[Kulsrud(2004)]{kulsrud_04} Kulsrud, R.~M. 2004, Plasma physics for astrophysics (Princeton University Press)
\bibitem[Maron \& Goldreich(2001)]{maron_g01} Maron, J., \& Goldreich, P. 2001, \apj, 554, 1175
   \bibitem[Mason et al.(2006)]{mason_cb06}
  Mason, J., Cattaneo, F. \& Boldyrev, S. 2006, \prl, 97, 255002
  \bibitem[Mason et al.(2008)]{mason_cb08}
  Mason, J., Cattaneo, F. \& Boldyrev, S. 2008, \pre, 77, 036403
\bibitem[McKee \& Ostriker(2007)]{mckee_o07} McKee, C.~F. \& Ostriker, E.~C., 2007, Ann. Rev. Astron. Astrophys., 45, 565 
\bibitem[M\"uller \& Grappin(2005)]{muller_g05} M\"uller, W.-C. \& Grappin, R. 2005, \prl, 95, 114502
\bibitem[Oughton, Dmitruk \& Matthaeus(2004)]{oughton_dm04} Oughton, S., Dmitruk, P. \& Matthaeus, W. H. 2004, Physics of Plasmas, 11, 2214
\bibitem[Perez \& Boldyrev(2008)]{perez_b08} Perez, J.~C. \& Boldyrev, S. 2008, Astrophys. J., 672, L61
\bibitem[Perez \& Boldyrev(2009)]{perez_b09} Perez, J. C. \& Boldyrev, S. 2009, \prl, 102, 025003
\bibitem[Perez \& Boldyrev(2010)]{perez_b10} Perez, J. C. \& Boldyrev, S. 2010, Phys. Plasmas, 17, 055903
 \bibitem[Perez et al.(2011)]{perez_etalp} Perez, J.C., Mason, J., Boldyrev, S. \& Cattaneo, F. 2011, in preparation
\bibitem[Podesta \& Borovsky(2010)]{podesta_b10} Podesta, J. J. \& Borovsky, J. E. 2010, Phys. Plasmas, 17, 112905 
\bibitem[Podesta et al.(2009)]{podesta_etal09}	 Podesta, J. J., Chandran, B. D. G., Bhattacharjee, A., Roberts, D. A., \& Goldstein, M. L. 2009, J. Geophys. Res., 114, A01107 
\bibitem[Politano \& Pouquet(1998)]{politano_p98b} Politano, H. \& Pouquet, A. 1998, Geophys. Res. Lett, 25, 273 
\bibitem[Schekochihin \& Cowley(2007)]{schekochihin_c07} Schekochihin, A.~A. \& Cowley, S.~C. 2007, in Magnetohydrodynamics: Historical Evolution and Trends, (Springer, Dordrecht), 85 
\bibitem[Strauss(1976)]{strauss_76} Strauss, H. 1976, Phys. Fluids, 19, 134
\end{thebibliography}
\end{document}